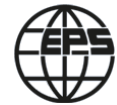

PAPER • OPEN ACCESS

# Do students benefit from drawing productive diagrams themselves while solving introductory physics problems? The case of two electrostatics problems

View the article online for updates and enhancements.

## Related content

- Investigating and improving introductory physics students' understanding of the electric field and superposition principle
  Jing Li and Chandralekha Singh

- Surveying college introductory physics students attitudes and approaches to problem solving
  Andrew J Mason and Chandralekha Singh

- Developing and validating a conceptual survey to assess introductory physics students' understanding of magnetism
  Jing Li and Chandralekha Singh

# Do students benefit from drawing productive diagrams themselves while solving introductory physics problems? The case of two electrostatics problems

Alexandru Maries[1] and Chandralekha Singh[2]

[1] Department of Physics, University of Cincinnati, Cincinnati, OH 45221, United States of America
[2] Department of Physics and Astronomy, University of Pittsburgh, Pittsburgh, PA 15260, United States of America

**Abstract**
An appropriate diagram is a required element of a solution building process in physics problem solving and it can transform a given problem into a representation that is easier to exploit for solving the problem. A major focus while helping introductory physics students learn problem solving is to help them appreciate that drawing diagrams facilitates problem solving. We conducted an investigation in which two different interventions were implemented during recitation quizzes throughout the semester in a large enrolment, algebra-based introductory physics course. Students were either (1) asked to solve problems in which the diagrams were drawn for them or (2) explicitly told to draw a diagram. A comparison group was not given any instruction regarding diagrams. We developed a rubric to score the problem solving performance of students in different intervention groups. We investigated two problems involving electric field and electric force and found that students who drew productive diagrams were more successful problem solvers and that a higher level of relevant detail in a student's diagram corresponded to a better score. We also conducted think-aloud interviews with nine students who were at the time taking an equivalent introductory algebra-based physics course in order to gain insight into how drawing diagrams affects the problem solving process. These interviews supported some of the interpretations of the quantitative results. We end by discussing instructional implications of the findings.

## 1. Introduction

Introductory physics is a challenging subject to learn. It is difficult for introductory students to associate the abstract concepts they study in physics with more concrete representations that facilitate understanding without an explicit instructional strategy aimed to aid them in this regard. Without guidance, introductory students often employ formula oriented problem solving strategies instead of developing a solid grasp of physical principles and concepts. There are many reasons to believe that multiple representations of concepts along with the ability to construct, interpret and transform between different representations that correspond to the same physical system or process play a positive role in learning physics. First, physics experts often use multiple representations as a first step in a problem solving process [1–3]. Second, students who are taught explicit problem solving strategies emphasising use of different representations of knowledge at various stages of problem solving construct higher quality and more effective representations and perform better than students who learn traditional problem solving strategies [4]. Third, multiple representations are very useful in translating the initial, usually verbal description of a problem into a representation more suitable to further analysis and mathematical manipulation [5, 6] partly because the process of constructing an effective representation of a problem makes it easier to generate appropriate decisions about the solution process. Also, getting students to represent a problem in different ways helps shift their focus from merely manipulating equations toward understanding physics [7]. Some researchers have argued that in order to understand a physical concept thoroughly, one needs to be able to recognise and manipulate the concept in a variety of representations [5, 7]. As Meltzer puts it [8], a range of diverse representations is required to 'span' the conceptual space associated with an idea. Since traditional courses, which generally do not emphasise multiple representations, lead to low gains on the Force Concept Inventory [9, 10] and on other assessments in the domain of electricity and magnetism [11, 12], in order to improve students' understanding of physics concepts, many researchers have developed instructional strategies that place explicit emphasis on multiple representations [1, 5, 13, 14] while other researchers developed other strategies with implicit focus on multiple representations [6, 15–19]. Van Heuvelen's approach [5], for example, starts by ensuring that students explore the qualitative nature of concepts by using a variety of representations of a concept in a familiar setting before adding the complexities of mathematics.

One representation useful in the initial conceptual analysis and planning stages of a solution is a schematic diagram of the physical situation presented in the problem. Diagrammatic representations have been shown to be superior to exclusively employing verbal or mathematical representations when solving problems [3, 20–22]. It is therefore not surprising that physics experts automatically employ diagrams in attempting to solve problems [1, 7, 23]. However, introductory physics students need explicit help to (1) understand that drawing a diagram is an important step in organising and simplifying the given information into a representation which is more suitable to further analysis [24], and (2) learn how to draw appropriate and useful diagrams. Therefore, many researchers who have developed strategies for teaching students effective problem solving skills use scaffolding support designed to help students recognise how important the step of drawing a diagram is in solving physics problems, and guidance to help them draw useful diagrams. In Newtonian mechanics, Reif [1] has suggested that several diagrams be drawn: one diagram of the problem situation which includes all objects and one diagram for each system that needs to be considered separately. Also, he described in detail concrete steps that students need to take in order to draw these diagrams as follows:

(a) describe both motions and interactions,
(b) identify interacting objects before forces,
(c) separate long range and contact interactions, and
(d) label contact points by the magnitude of the action–reaction pair of forces.

Van Heuvelen's Active Learning Problem Sheets [5] adapted from Reif follow a very similar underlying approach. Other researchers who have emphasised, among other things, the importance of diagrams in their approach to teaching students problem solving skills have found significant improvements in students' problem solving methods [2, 5, 25].

Previous research shows that students who draw diagrams even if they are not rewarded for it are more successful problem solvers [17]. In addition, students who take courses which emphasise effective problem solving heuristics which include drawing a diagram are more likely to draw diagrams even on multiple-choice exams [25]. An investigation into how spontaneous drawing of free body diagrams (FBDs) [26] affects problem solving [27] shows that only drawing correct FBDs improves a student's score and that students who draw incorrect FBDs do not perform better than students who draw no diagrams. Heckler [28] investigated the effects of prompting students to draw FBDs in introductory mechanics by explicitly asking students to draw clearly labelled FBDs. He found that students who were prompted to draw FBDs were more likely to follow formally taught problem solving methods rather than intuitive methods which sometimes caused deteriorated performance.

This study is part of a larger investigation on the impact of using multiple representations in physics problem solving [29], and one of the principal types of representations investigated were diagrammatic representations. The broad questions in this larger investigation as pertaining to diagrammatic representations were:

1. Is there a correlation between drawing diagrams and problem solving performance even when (i) students are not graded on drawing diagrams, and (ii) the solution to the problem involves primarily mathematical manipulation of equations?
2. Should students be provided diagrams or asked to draw them while solving introductory physics problems?

The larger investigation also explored mathematical and graphical representations [30], in particular, the extent to which students' mathematical and graphical representations of electric field were consistent with one another and the impact of two different scaffolding supports designed to help students make the connection between the two representations. In this study, we primarily explored question 1.(i), although the insight gained from this study can be used to inform question 2. We investigated how prompting students to draw diagrams (without being more specific, e.g. prompting students to draw FBDs as in [28]) affects their performance in two electrostatics problems and how their performance is impacted when provided with a diagrammatic representation of the physical situation described in the problems. There has been much research on students' conceptual understanding of electricity and magnetism [31–40]. However, here, we are primarily interested in how students who draw diagrams themselves as part of the problem solving process benefit (in their problem solving performance) from their diagrams and we also investigate how diagrams drawn by students, classified as productive or unproductive based upon certain criteria, affect student performance. In addition to the quantitative data collected, we conducted think-aloud interviews [41] with nine students who were taking an equivalent introductory algebra-based physics course at the time, to gain insight into how drawing (or not drawing) diagrams may affect their problem solving performance. The interviews provided possible interpretations for some of the quantitative findings.

## 2. Methodology

### 2.1. In-class study

A traditionally taught class of 120 algebra-based introductory physics students was enroled in three different recitation sections. The three recitation sections formed the comparison group and the two intervention groups for this investigation. All recitations were taught traditionally: the TA worked out problems similar to the homework problems and then gave a 15 min quiz at the end of class. Students in all recitations attended the same lectures, were assigned the same homework, and took the same exams and quizzes. In the recitation quizzes throughout the semester, students in the three different recitation sections were given the same problems but with the following interventions:

(1) *Prompt only group (PO)*: in each quiz problem, students were given explicit instructions to draw a diagram with the problem statement;
(2) *Diagram only group (DO)*: in each quiz problem, students were provided a diagram drawn by the instructor intended as scaffolding support to aid in solving the problem; and
(3) *No support group (NS)*: this group, the comparison group, was not given any diagram or explicit instruction to draw a diagram with the problem statement.

We note that students in the DO group were provided with copies of the quiz problems which had the diagrams drawn. Some students annotated the provided diagrams by adding relevant details, other students drew their own diagrams, and yet other students did not annotate the provided diagrams but did not draw their own diagrams either.

The sizes of the different recitation groups varied from 22 to 55 students because the students were not assigned a particular recitation; they could choose to attend any of the three each week. For the same reason, the size of each recitation group also varied from week to week, although not as drastically because most students (≈80%) would stick with a particular recitation. Furthermore, each intervention was not matched to a particular recitation. For example, in one week, students in the Tuesday recitation comprised the comparison group, while in another week the comparison group was a different recitation section. This is important because it implies that individual students were subjected to different interventions from week to week so that we do not expect cumulative effects due to the same group of students always being part of the same intervention.

In this paper, we analyse two problems: the first problem is one-dimensional and has two almost identical parts, one dealing with electric field and the other dealing with electric force. This problem was given both in a quiz (a week after students learned about these concepts) and in a midterm exam (several weeks after learning the concepts). Note that the interventions were only implemented in the quiz and not in the midterm. Also, students received feedback from the TAs about their performance on the quiz (i.e. the TAs graded student solutions, marked mistakes and returned the quizzes). Solutions to the quizzes were also provided to the students before the midterm exam. The second problem is a two-dimensional problem on electric force which was given in a quiz only. The two problems and the diagrams provided to students in the DO group (shown in figures 1 and 2) are the following:

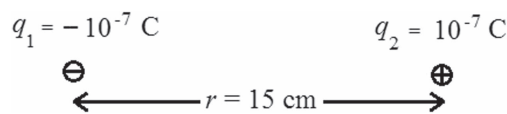

**Figure 1.** Diagram for problem 1 given only to students in the DO group.

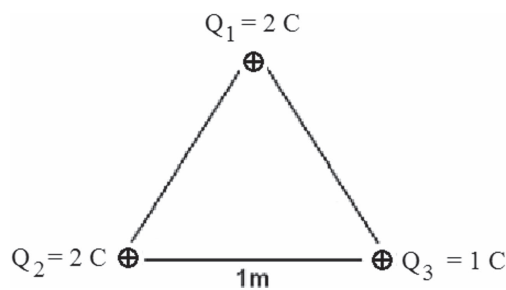

**Figure 2.** Diagram for problem 2 given only to students in the DO group.

**Problem 1**

'Two equal and opposite charges with magnitude $10^{-7}$ C are held 15 cm apart.

(a) What are the magnitude and direction of the electric field at the point midway between the charges?
(b) What are the magnitude and direction of the force that would act on a $10^{-6}$ C charge if it is placed at that midpoint?'

**Problem 2**

'Three charges are located at the vertices of an equilateral triangle that is 1 m on a side. Two of the charges are 2 C each and the third charge is 1 C. Find the magnitude and direction of the net electrostatic force on the 1 C charge'.

These diagrams were drawn by the instructor. They are very similar to what most physics experts would draw in the initial stage of problem solving. Of course, subsequently, physics experts would most likely augment these diagrams by drawing arrows to indicate the directions of electric field/force vectors. Providing students in the DO group with these diagrams was intended as a scaffolding support based upon the hypothesis that the pictorial representation of a problem situation can aid students in visualising the problem.

In order to ensure homogeneity of grading, we developed rubrics for each problem analysed and ensured that there was at least 90% inter-rater reliability between two different raters on at least 10% of the data. The development of the rubric for each problem went through an iterative process. During the development of the rubric, the two graders also discussed a student's score separately from the one obtained using the rubric and adjusted the rubric if it was agreed that the version of the rubric was too stringent or too generous. After each adjustment, all students were graded again on the revised rubric. Problem 1 is comprised of two subproblems, part (a) which asks for electric field and part (b) which asks for electric force. Therefore, parts (a) and (b) were scored separately. In table 1, we provide the summary of the rubric for part (a) (electric field) of the first problem. The rubric for part (b) (electric force) is very similar. Student performance on problem 2 was scored in a similar manner (used a rubric developed through an iterative process and ensured 90% inter-rater reliability between two different raters on at least 10% of the data).

**Table 1.** Summary of the used rubric for part (a) of problem 1.

| | Correct ideas | |
|---|---|---|
| Section 1 | Used correct equation for the electric field | 1p |
| Section 2 | Added the two fields due to individual charges correctly | 7p |
| Section 3 | Indicated correct direction for the net electric field | 1p |
| Section 4 | Correct units | 1p |
| | Incorrect ideas | |
| Section 1 | Used incorrect equation for the electric field | −1p |
| Section 2 | 2.1 Did nothing in this section | −7p |
| | 2.2 Did not find electric fields due to both charges | −6p |
| | 2.3 Used Pythagorean theorem (not relevant here) or obtained zero for electric field | −4p |
| | 2.4 Did not use $r/2$ to find the electric field | −2p |
| | 2.5 Minor mistake(s) in finding the electric field | −1p |
| Section 3 | Incorrect or no mention of the direction of the net electric field | −1p |
| Section 4 | Incorrect or no units | −1p |

Table 1 shows that there are two parts to the rubric: correct and incorrect ideas. Table 1 also shows that in the correct ideas part, the problem was divided into different sections and points were assigned to each section. Each student starts out with 10 points and in the incorrect ideas part we list the common mistakes students made in each section and how many points were deducted for each of those mistakes. We note that it is not possible to deduct more points than a section is worth (the mistakes labelled 2.1 and 2.2 are exclusive with respect to all other mistakes in section 2 and with each other). We also left ourselves a small window (labelled 2.5) if the mistake a student made was not explicitly in the rubric (only 5% of the cases).

For example, one common mistake on problem 1 was to use the equation for the magnitude of the electric field of a point charge $\left(|\vec{E}| = \frac{kQ}{r^2}\right)$, plug in the magnitude of one charge ($10^{-7}$ C) and the distance between them (15 cm) to obtain $4 \times 10^4$ N C$^{-1}$. Some students who did this drew arrows which indicate that the charges are attracted to each other (i.e. a rightward force on the negative charge and a leftward force on the positive charge), and did not indicate the direction of the net electric field. Here is how the rubric in table 1 was applied to this type of student solution:

- Section 1: 1 point since students used the correct equation
- Section 2: 1 point because 2.2 is the mistake students made (also, as mentioned above, mistake 2.2 was considered to be exclusive with all other mistakes in section 2)
- Section 3: no points because the direction of the net field is not indicated
- Section 4: 1 point because units are correct
- Total: 3/10 points.

## 2.2. Out-of-class study: think-aloud interviews

In addition to the quantitative in-class data collected, individual interviews were conducted using a think-aloud protocol [41] with nine students who were at the time enroled in a second semester algebra-based introductory physics course. During the interviews, students were asked to solve the problems while thinking aloud and, after they had finished working on the problems, they were asked follow-up questions related to the physics concepts required for successfully solving the problems. The interviews provided qualitative data which provided an interpretation for some of the quantitative findings.

## 3. Research questions

Below, we discuss the research questions investigated in this study. The first two are specific to the interventions and the other two are more general and related to the effect of drawing a diagram on problem solving performance.

*RQ1: How do the different interventions affect the frequency of students drawing productive diagrams?*

Physics experts would most likely augment the diagrams provided by drawing arrows which represent the directions of electric field/electric force vectors. Therefore, it was considered that a productive diagram should include at the very least, in addition to the charges, two electric field or electric force vectors (for example, for problem 1, two vectors which indicate the direction for electric fields/electric forces explicitly drawn at the midpoint, whether or not another charge is placed there. For problem 2, two vectors which indicate the directions of the two forces which act on the 1 C charge). Any diagram which did not include vectors to indicate directions of electric fields and/or forces was considered to be unproductive. Productive diagrams can include more relevant detail. For example, in problem 2, in addition to the two forces that act on the 1 C charge, a student can explicitly draw the components of those forces. It is worthwhile noting that for both problems, students in the DO group were provided unproductive diagrams.

*RQ2: To what extent is student performance influenced, if at all, by the interventions?*

Since the first step of most physics experts in problem solving is conceptual planning and analysis, which typically includes drawing one or several diagrams, it is possible that prompting students to draw diagrams can make it more likely that they engage in this planning stage, which may help their problem solving performance. Providing a diagram might also affect their performance. We investigated how students in the two different intervention groups performed compared to the students in the comparison group.

*RQ3: To what extent does drawing a productive diagram affect problem solving performance?*

In a previous investigation [22], we found that students who drew productive diagrams performed better than students who did not draw a productive diagram for a problem involving a standing harmonic of a sound wave in a cylindrical tube. We investigated whether this effect also arises in the context of the problems discussed here.

*RQ4: What are some possible cognitive mechanisms that can explain the effect of drawing a productive diagram on student performance?*

In order to shed light on possible cognitive mechanisms which could partly explain how students' problem solving performance is affected (or not affected) by drawing a diagram, nine think-aloud interviews were conducted with students enroled in a different, but equivalent algebra-based introductory physics course. At the time of the interviews, students had finished the study of electrostatics and also had been tested on this material via an in-class exam.

**Table 2.** Percentages (and numbers) of students in the three intervention groups who drew a productive diagram for problem 1.

|  | % of students who drew a productive diagram (number of students who drew a productive diagram) |
| --- | --- |
| PO | 66% (19) |
| DO | 45% (18) |
| NS | 41% (21) |
| All students | 50% (58) |

**Table 3.** Percentages (and numbers) of students in the three intervention groups who drew a productive diagram for problem 2.

|  | % of students who drew a productive diagram (number of students who drew a productive diagram) |
| --- | --- |
| PO | 82% (41) |
| DO | 79% (31) |
| NS | 66% (19) |
| All students | 77% (91) |

## 4. Results

### 4.1. RQ1: How do the different interventions affect the frequency of students drawing productive diagrams?

For both problems, all students drew a diagram. However, not all diagrams drawn by students were considered to be productive (for the purposes of solving the problems). In problem 1, intervention PO resulted in significantly increasing the percentage of students who drew a productive diagram ($p$ value = 0.036 when compared to NS via a chi-square test [42]) while the percentage of students in DO who drew a productive diagram is nearly identical to the percentage of students in NS, as shown in table 2. Note that since students in the DO group were provided with an unproductive diagram, only 45% of them added more detail to those diagrams to obtain a productive diagram. For problem 2, neither intervention affected the percentage of students who drew productive diagrams significantly (data shown in table 3). We note however, that problem 2 is two-dimensional while problem 1 is one-dimensional and that more students drew productive diagrams for problem 2 than for problem 1 (77% compared to 50%).

### 4.2. RQ2: To what extent is student performance influenced, if at all, by the interventions?

Similar to the percentage of students who drew a productive diagram discussed above, it appears that while the interventions had some effect on student performance for problem 1, they did not have an effect for problem 2. Table 4 lists the average score for each group (PO, DO, NS) on the two different parts for problem 1 (given in a quiz, one week after students learned about electric field and electric force). ANOVA [42] indicates no statistically significant difference between the three groups on the electric field part ($p = 0.332$), but on the electric force part, the three groups are not all comparable in terms of performance ($p = 0.040$). In order to investigate further, pair-wise $t$-tests [42] were carried out for the electric force part which indicate that students in the PO group performed significantly better than students in the two other groups (comparing PO with DO: $p$ value $= 0.017$, effect size $= 0.60$; comparing PO with NS: $p$ value $= 0.011$, effect size $= 0.55$). These effect sizes correspond to medium effects.

**Table 4.** Number of students ($N$), averages (Avg.) and standard deviations (S.d.) on the two parts of problem 1 for the two intervention groups and the comparison group out of 10 points.

|    | Field part | | | Force part | |
|----|---|------|------|------|------|
|    | $N$ | Avg. | S.d. | Avg. | S.d. |
| PO | 29 | 7.0 | 3.25 | 8.6 | 2.80 |
| DO | 40 | 7.1 | 2.61 | 6.6 | 3.77 |
| NS | 51 | 7.9 | 2.78 | 6.8 | 3.59 |

**Table 5.** Number of students ($N$), averages and standard deviations (Std. dev.) on problem 2 for the two intervention groups and the comparison group out of 10 points.

|    | $N$ | Average | Std. dev. |
|----|-----|---------|-----------|
| PO | 50 | 5.8 | 3.1 |
| DO | 39 | 6.7 | 2.5 |
| NS | 29 | 5.3 | 3.3 |

On problem 2, ANOVA indicated no statistically significant differences between the different groups ($p = 0.131$), possibly because on problem 2, the percentages of students who drew a productive diagram in the three different groups were comparable. The averages and standard deviations of students in the three different groups are shown in table 5 (the sizes of the intervention groups in tables 4 and 5 do not match because the two problems investigated here were given in two different quizzes and the interventions were implemented in different recitations in different weeks).

It therefore appears that for problem 1, students who were asked to draw a diagram performed significantly better (in the force part of the problem at least), perhaps because they were more likely to draw productive diagrams, while for problem 2, the interventions did not show significantly different trends (percentage of students drawing a productive diagram or score).

### 4.3. RQ3: To what extent does drawing a productive diagram affect problem solving performance?

<u>Students who draw productive diagrams perform better than students who do not</u>

As mentioned earlier, productive diagrams for both problems include the basic physical setups (i.e. two charges from problem 1) and vectors which indicate the directions of electric field or electric force vectors. Table 6 shows the performance of students who drew productive diagrams and those who did not for both problems regardless of the intervention (i.e. all students are put together). Our results indicate that students who drew a productive diagram significantly outperformed students who did not on both problems (both $p$ values are less than 0.001 and effect sizes correspond to large effects), which is similar to a result previously encountered in the context of students' problem solving performance on a problem involving standing sound waves in tubes [22].

**Table 6.** Number of students (*N*), averages and standard deviations (Std. dev.) for students who drew productive diagrams and those who did not on problems 1 and 2 out of 10 points, and *p* values and effect sizes for comparing the performance of students who drew a productive diagram with the performance of students who did not draw a productive diagram.

|  | N | Average | Std. dev. | *p* value | Effect size |
| --- | --- | --- | --- | --- | --- |
| Problem 1—drew prod. diag. | 58 | 8.3 | 2.2 | < 0.001 | 0.84 |
| Problem 1—did not draw prod. diag. | 62 | 6.3 | 2.6 | | |
| Problem 2—drew prod. diag. | 91 | 6.6 | 2.9 | < 0.001 | 0.91 |
| Problem 2—did not draw prod. diag. | 27 | 4.1 | 2.5 | | |

<u>A higher level of relevant detail in a student's diagram corresponds to better performance</u>

For both problems 1 and 2, students drew productive diagrams which included varying levels of relevant detail. For example, for problem 1, some students drew the two charges as well as two electric field vectors at the midpoint (relevant detail 1). Other students drew the two charges, and also drew two electric field and two electric force vectors at the midpoint (relevant detail 2). Typically, students who drew the latter type of diagram had two separate diagrams, one for the electric field part and one for the electric force part. And yet other students drew an unproductive diagram which does not include vectors indicating directions of electric field or force vectors at the midpoint. We note that students may also add details to the diagrams that are not directly relevant for the problem solving process. For example, some students drew vectors going outward from the positive charge and inward towards the negative charge in all directions. Interviews suggest that some of these students were replicating what the electric field looks like around isolated positive and negative charges. While this is related to the physical situation presented in the problem, it is not directly useful for solving the problem unless the students recognise that they need to consider the midpoint and think about the direction of the electric field caused by each charge at that point. Thus, the researchers considered that a productive diagram must have detail that is directly relevant to solving the problem.

For problem 2, productive diagrams included, e.g. the three charges and the two forces acting on the 1 C charge (relevant detail 1), or the three charges, the two forces acting on the 1 C charge and their *x* and *y* components drawn for a particular choice of coordinate system (relevant detail 2). An unproductive diagram included only the three charges. Similar to problem 1, some students added details to their diagram that were not directly relevant for solving the problem. For example, some students drew vectors indicating the direction of the forces acting on the two 2 C charges. Those details may be useful if the problem asks for the net forces acting on the 2 C charges, but they are not useful for finding the net force on the 1 C charge. Thus, for both problems 1 and 2, our consideration of what features of a diagram make it productive relates to visualising relevant information from the problems that is useful for solving them (e.g. directions of electric field or electric force vectors).

Table 7, which shows the performance of students who drew these different types of diagrams for both problems, indicates that a higher level of relevant detail in a student's diagram corresponds to a higher score. For problem 1 (1D problem), which was given both in a quiz and in a midterm, there is a statistically significant difference between students who

**Table 7.** Numbers of students ($N$), averages (Avg.) and standard deviations (Std. dev.) for groups of students with diagrams including different levels of relevant detail for problem 1 in the quiz and the midterm, and for problem 2 in the quiz.

|  | Problem 1—Quiz | | | Problem 1—Midterm | | |
| --- | --- | --- | --- | --- | --- | --- |
|  | $N$ | Avg. | Std. dev. | $N$ | Avg. | Std. dev. |
| Unproductive diagram | 62 | 6.4 | 2.6 | 45 | 7.0 | 2.6 |
| Relevant detail 1 | 49 | 8.3 | 2.2 | 51 | 8.4 | 2.0 |
| Relevant detail 2 | 9 | 8.9 | 1.4 | 25 | 9.0 | 1.4 |

|  | Problem 2—Quiz | | |
| --- | --- | --- | --- |
|  | $N$ | Avg. | Std. dev. |
| Unproductive diagram | 27 | 4.1 | 2.5 |
| Relevant detail 1 | 58 | 5.7 | 2.9 |
| Relevant detail 2 | 33 | 8.0 | 2.2 |

**Table 8.** $p$ values and effect sizes for comparison of the performance of students with diagrams including different levels of relevant detail (UD = unproductive diagram, RD1 = relevant detail 1, RD2 = relevant detail 2) for problem 1 in the quiz and in the midterm and for problem 2 (in the quiz).

|  | UD-RD1 | | RD1-RD2 | |
| --- | --- | --- | --- | --- |
|  | $p$ value | Effect size | $p$ value | Effect size |
| Problem 1—Quiz | <0.001 | 0.82 | 0.284 | 0.33 |
| Problem 1—Midterm | 0.003 | 0.62 | 0.133 | 0.35 |
| Problem 2 | 0.008 | 0.61 | <0.001 | 0.87 |

drew unproductive diagrams and students who drew diagrams which included more relevant detail (both $p$ values for comparing students who drew relevant detail 1 or 2 diagrams with students who drew unproductive diagrams are less than 0.001, and the effect sizes are large), but the difference in performance between students who drew relevant detail 1 diagrams and students who drew relevant detail 2 diagrams is not statistically significant, as shown in table 8. This was found both in the quiz and in the midterm. For problem 2 (2D problem), students who drew relevant detail 1 diagrams performed significantly better than students who drew unproductive diagrams ($p = 0.008$, effect size $= 0.61$) and students who drew relevant detail 2 diagrams performed significantly better than students who drew relevant detail 1 diagrams ($p < 0.001$, effect size $= 0.87$). The differences between the averages of the groups are quite noticeable and the effect sizes point to medium to large effects despite the large variation within each group. The performance of students who drew diagrams with the highest level of relevant detail is nearly twice that of students who drew unproductive diagrams!

### 4.4. RQ4: What are some possible cognitive mechanisms that can explain the effect of drawing a productive diagram on student performance?

As mentioned earlier, individual interviews with nine students who were at the time taking an equivalent second semester of an introductory algebra-based physics course were carried out using a think-aloud protocol [41]. These interviews suggested that for problem 2, cognitive load theory [43] may be one possible framework that can explain why students who explicitly drew the components of the two forces performed better. In particular, two of the students interviewed were almost identical in terms of their majors and grades (both in the current physics course and the previous one). Karen and Dan were both Biology majors and in the first semester of physics they obtained similar grades (B+ and A−, respectively). In the second semester physics class, on the first exam (class average 75/100), they both obtained 81/100 and on the second exam (class average 65/100) they also both obtained 81/100. Note that the first exam was focused primarily on electrostatics and included questions which asked students to calculate the net electric field due to a configuration of charges and the net force acting on a particular charge, but the questions were in other contexts.

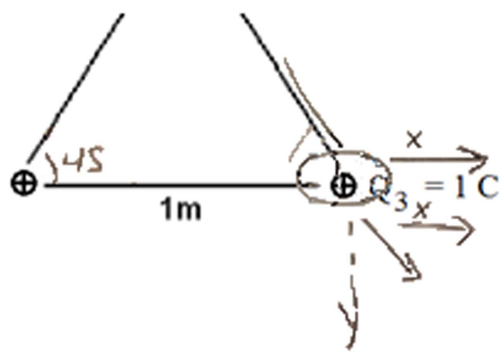

**Figure 3.** Forces due to the two individual charges on the 1 C charge and their components as drawn by Karen (student).

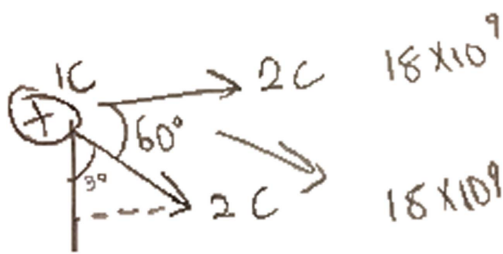

**Figure 4.** Forces acting on the 1 C charge due to the two 2 C charges as drawn by Dan (student).

When solving problem 2, Karen recognised that she needed to find the $x$ and $y$ components of both forces due to each of the 2 C charges and, before she proceeded to find them, she drew all the components on the diagram provided as shown in figure 3. She then figured out all the components and combined them correctly to determine both the magnitude of the net force and its direction (angle below the $x$ axis). While working on this problem, it was evident that Karen was focusing on only a few things at a time and was being systematic about the way in which she found the net force. For example, when finding the components of the oblique (not horizontal) force, she redrew a triangle in which this force was the hypotenuse and identified the angles. Karen's only mistake was that she used an angle of 45° instead of 60° to find these components.

Dan also immediately recognised that components should be considered and proceeded to find them after redrawing the 1 C charge (see figure 4) and the two forces acting on it due to the two 2 C charges. He worked more slowly than Karen on this problem, but after some time, he correctly determined the *x* and *y* components of the oblique force and wrote them down (trigonometric functions were still included, i.e. he wrote down the *y* component as $18 \times 10^9$ cos 30). However, unlike Karen, he did not draw these components on his diagram; his diagram of the forces (shown in figure 4) only included the two forces and their magnitudes.

When Dan combined the components, he made two mistakes: (1) his net *y* component did not include the trigonometric function which he had previously written down (when he found the *y* component of the oblique force). As he was determining the net *y* component he said: 'this one [horizontal force] doesn't have a *y* component, so it [the *y* component of the net force] is just 18 times $10^9$ [magnitude he found for the oblique force]' and (2) he subtracted the two *x* components instead of adding them (he subtracted the horizontal force from the *x* component of the oblique force). In particular, he wrote the following on the paper for the net *x* component:

$$\text{Net } x = 18 \times 10^9 \sin 30 - 18 \times 10^9.$$

It is possible that part of the reason why he subtracted the components is because he did not explicitly draw the *x* component of the oblique force and perhaps, due to the fact that the oblique force is in the fourth quadrant (which should be dealt with carefully), he implicitly assumed that one of its components must be negative, or that something must be subtracted. He subtracted the horizontal force from the *x* component of the oblique force even though the picture he drew clearly indicated that the horizontal force is in the positive *x* direction. After he finished working on all the problems to the best of his ability, in the second phase of the interview, he was asked for clarifications of points he had not made clear earlier and some additional questions. For example, Dan was asked a simpler question. He was asked to add two forces: one in the positive *y* direction, the other in the first quadrant, making an angle of 30° with the horizontal. Here too, he did not draw the components explicitly in the diagram and ended up subtracting the *y* components of the two forces in exactly the same manner in which he subtracted the *x* components in problem 2 (the triangle problem) i.e. he subtracted the vertical force from the *y* component of the oblique force. When asked why he subtracted these components he looked at the diagram for a few seconds and said:

> *Actually, you're adding […] sorry, I don't know why [I did that] […], you're adding because there's a positive y component here [vertical force] and a positive y component here [of the oblique force].*

The approaches of these two students differed mainly in that Karen explicitly drew all forces and components, whereas Dan only drew the forces. Dan subtracted the *x* components without providing a reason, and when he was asked to add two forces in a mathematical context (similar to the two forces in the physics context), he made exactly the same mistake for the two components that were supposed to be added. When questioned about why he subtracted them, he realised this mistake on his own almost immediately, which suggested that when he solved both problems (problem 2 and the simpler mathematical problem which had similar addition of vectors) he was not focusing on the appropriate information. Once his attention was drawn to the issue of whether the vectors should be added or subtracted in the simpler mathematical problem, he clearly knew that the *y* components must be added. Before questioning, he did not draw the components of the oblique force and appeared to be subtracting the components automatically, without a clear reason. Also, when asked why he subtracted the components, he did not start by trying to justify this (for example by beginning a sentence with 'I subtracted them because…'), which suggested that there was no clear reason for why he subtracted the *y* components. In other words, it is possible that he did not have any cognitive resources free to use for thinking about whether the components should be added or subtracted due to having to process too much information at one time in his working memory (e.g. forces, trigonometric angles, vector addition, etc). When it was time to utilise this information about the components of the oblique force to find the $x$ component of the net force, he forgot to correctly account for the $x$ component. On the other hand, Karen had the components explicitly drawn on the paper as opposed to keeping this information in her head and she was able to look back at her components and account for the sign of the $x$ component of the oblique force correctly. Cognitive load theory [43], which incorporates the notion of distributed cognition [44, 45], provides one possible explanation for Dan's unsuccessful and Karen's successful addition of vectors in this context: lack of information about components on Dan's diagram required him to keep this information in his working memory, while Karen did not need to keep this information in her working memory since she included the components explicitly in her diagram. As Dan's working memory was processing a variety of information during problem solving, he may have had cognitive overload and the information about the components that he planned to use at the opportune time to find the components of the net force was not invoked appropriately.

Interviews with other students who drew diagrams which included higher levels of detail suggested that including information on a diagram can help free up cognitive resources for processing information about vector addition and about the problems in general which helped them perform appropriate calculations and find their mistakes. On the other hand, students who drew unproductive diagrams or no diagrams at all sometimes seemed to have cognitive overload since, similar to Dan, they made mistakes while solving the problems initially. However, when coming back to the problem after being asked about their solutions, they sometimes identified their mistakes on their own. This suggested that when they solved the problems initially they may not have carefully carried out decision making regarding the problem solution partly because they had reduced cognitive processing capacity. Including information about the problems explicitly, e.g. by using diagrammatic representations can help increase the students' cognitive processing capacity by distributing their cognition [44, 45].

## 5. Discussion and summary

We found that for problem 1, students who were explicitly asked to draw a diagram were more likely to draw a productive diagram. We also found that students who drew productive diagrams performed better than students who drew unproductive diagrams. Among the students provided with a diagram (which was unproductive unless modified by the student by adding force and/or field vectors at the midpoint), less than half added relevant details to the diagram provided in order to use a productive diagram. This is a statistically significantly lower percentage compared to the percentage of students who used a productive diagram in the group of students who were prompted to draw one. This finding suggests that in an introductory physics course, prompting students to draw a diagram may provide better scaffolding for solving problems than providing a diagram and should be incorporated in helping students learn effective problem solving strategies. Furthermore, we also found that diagrams which included more relevant details from the problems (that are useful for solving the problems) corresponded to better performance. This finding suggests that students should not only be incentivized to draw diagrams, but also guided to learn to include as much relevant information as is necessary in their diagrams to facilitate problem solution. As noted earlier, one theoretical framework that can provide a possible explanation for why students with diagrams with more relevant details performed better is the cognitive load theory [43], which incorporates the notion of distributed cognition [44, 45]. In problem 2, students had to add forces by using components, so students who did not draw the force vectors or their components they had to add vectorially would have to keep too much information in the working memory [46–48] while engaged in problem solving (individual components of the two forces, angles required to get those components, what trigonometric function needs to be used for each component, etc). This can lead to cognitive overload and deteriorated performance. Explicitly drawing the forces and their components can reduce the amount of information that must be kept in the working memory while engaged in problem solving and may therefore make the student better able to go through all the steps necessary without making mistakes.

It is also important to note that these problems were given in the second semester of a one year introductory physics course for algebra-based students. These students had done problems for which they had to find the net force in Newtonian mechanics, and still less than 30% of the students realised that they should draw the components of the electric force in problem 2 presented here. Also, only 42% of all students indicated a direction for the net force. This can partly be an indication of a lack of transfer from one context to another [49, 50]. Students' performance also suggests that many algebra-based introductory students do not have a robust knowledge structure of physics nor do they employ good problem solving heuristics and their familiarity with addition of vectors may also require an explicit review. Earlier surveys at the start of the course have found that only about 1/3 of the students in an introductory physics class of the type discussed here had sufficient knowledge about vectors to begin the study of Newtonian mechanics [51]. Here we find that even after a semester of instruction in physics that involves a fair amount of vector addition, the fraction remains about the same and students had great difficulty dealing with vector addition in component form.

This study suggests that students drawing and using productive diagrams can help improve their problem solving performance, and suggests multiple avenues for future research. For example, one can conduct a more detailed investigation of the features that constitute a productive or an unproductive diagram and how those are correlated with problem solving performance. We should note that our study suggests that the features of a productive diagram are related to representing relevant information from the problem to facilitate problem solving, but this could be explored in more detail in future studies. Also, while this study suggests that asking students to draw a diagram may provide useful scaffolding for students, one can also investigate other possible interventions, for example, providing students with a diagram and explicitly asking them to add details to it, or providing students with productive diagrams. Future studies could also explore possible reasons why some students draw diagrams while others do not, as well as the characteristics of teacher-student interactions that may help students recognise that they should not only draw diagrams but also ensure that relevant details are included in those diagrams. Our study suggests that representing relevant information from electrostatics problems on a diagram that students drew helped them in the problem solving process. Instructors may emphasise this in their teaching, as well as discuss that visual information is much easier to process than verbal information and this is partly why physics experts always draw diagrams when solving problems. While the extent to which such practices may be effective is beyond the scope of this work, future research can explore these issues along with other approaches (e.g. providing grade incentives for drawing diagrams).

## Acknowledgments

We thank the US National Science Foundation for award 1524575 which made this work possible. Also, we are extremely grateful to professors F Reif, J Levy and R P Devaty, and all the members of the Physics Education Research group at University of Pittsburgh for very helpful discussions and/or feedback.

## ORCID iDs

Alexandru Maries  https://orcid.org/0000-0001-5781-7343
Chandralekha Singh  https://orcid.org/0000-0002-1234-5458